\documentclass[reprint,twocolumn,superscriptaddress,nofootinbib,amsmath,amssymb]{aastex63}
\usepackage{graphicx}
\usepackage{dcolumn}
\usepackage{bm}
\usepackage{hyperref}
\hypersetup{
  colorlinks=true,        
  linkcolor=blue,         
  citecolor=cyan,         
  urlcolor=cyan            
}
\usepackage{colortbl}
\usepackage{amsmath}
\usepackage{times}

\newcommand{\ie}{i.e.,~}
\newcommand{\eg}{e.g.,~}

\begin{document}

\title[On the Sound Speed in Neutron Stars]{On the Sound Speed in Neutron Stars}

\author{Sinan Altiparmak}
\affiliation{Institut f\"ur Theoretische Physik, Goethe Universit\"at,
  Max-von-Laue-Str. 1, 60438 Frankfurt am Main, Germany}

\author[0000-0002-8669-4300]{Christian Ecker}
\affiliation{Institut f\"ur Theoretische Physik, Goethe Universit\"at,
  Max-von-Laue-Str. 1, 60438 Frankfurt am Main, Germany}

\author[0000-0002-1330-7103]{Luciano Rezzolla}
\affiliation{Institut f\"ur Theoretische Physik, Goethe Universit\"at,
  Max-von-Laue-Str. 1, 60438 Frankfurt am Main, Germany}
\affiliation{School of Mathematics, Trinity College, Dublin 2, Ireland}
\affiliation{Frankfurt Institute for Advanced Studies,
  Ruth-Moufang-Str. 1, 60438 Frankfurt am Main, Germany}
 
\begin{abstract}
Determining the sound speed $c_s$ in compact stars is an important open
question with numerous implications on the behaviour of matter at large
densities and hence on gravitational-wave emission from neutron stars. To
this scope, we construct more than $10^7$ equations of state (EOSs) with
continuous sound speed and build more than $10^8$ nonrotating stellar
models consistent not only with nuclear theory and perturbative QCD, but
also with astronomical observations. In this way, we find that EOSs with
sub-conformal sound speeds, \ie with $c^2_s < 1/3$ within the stars, are
possible in principle but very unlikely in practice, being only $0.03\%$
of our sample. Hence, it is natural to expect that $c^2_s > 1/3$
somewhere in the stellar interior. Using our large sample, we obtain
estimates at $95\%$ credibility of neutron-star radii for representative
stars with $1.4$ and $2.0$ solar masses,
$R_{1.4}=12.42^{+0.52}_{-0.99}\,{\rm km}$,
$R_{2.0}=12.12^{+1.11}_{-1.23}\,{\rm km}$, and for the binary tidal
deformability of the GW170817 event,
$\tilde\Lambda_{1.186}=485^{+225}_{-211}$. Interestingly, our
lower-bounds on the radii are in very good agreement with the prediction
derived from very different arguments, namely, the threshold
mass. Finally, we provide simple analytic expressions to determine the
minimum and maximum values of $\tilde\Lambda$ as a function of the chirp
mass.
\end{abstract}

\keywords{neutron stars, equation of state, sound speed}

\section{Introduction}

Gravity compresses matter in the core of a
neutron star to densities a few times larger than the saturation (number)
density of atomic nuclei $n_0:=0.16\,\rm fm^{-3}$.
A quantity that describes the stiffness of matter, a property required to
prevent a static neutron star from collapsing to a black hole, is the
sound speed
\begin{equation}
c_s^2:=\left(\frac{ \partial  p}{ \partial  e}\right)_s\,,
\end{equation}
where the pressure $p$ and the energy density $e$ are related by the
equation of state (EOS) and the derivative is considered at constant
specific entropy $s$. Calculating the EOS of cold matter at the densities
reached in the innermost part of a neutron star is still an open
problem. Causality and thermodynamic stability imply $0\leq c_s^2\leq 1$,
which poses the minimal requirement the EOS has to satisfy. Beyond this
basic causality requirement, there are theoretical constraints on the
EOS, and therefore on the sound speed, available in two different
regimes. In particular, at small densities, \ie $n\lesssim \,n_0$, the
EOS is well described by effective field theory models~\citep{Hebeler:2013nza, 
Gandolfi2019, Keller:2020qhx} and the corresponding
sound speed is found to be small $c_s^2\ll 1$. At larger densities --
much higher than those realised inside neutron stars, \ie $n\gg \,n_0$ --
matter is in a state of deconfined quarks and gluons and the EOS of
Quantum Chromodynamics (QCD) becomes accessible to perturbation
theory. Because conformal
symmetry of QCD is restored at asymptotically large densities, the sound
speed approaches the value in conformal field theory and realized in
ultrarelativistic fluids $c_{s}^2=1/3$ from below. Between these two limits, the
EOS is not accessible by first-principle techniques and the sound speed
is essentially unknown.

Given this considerable uncertainty, at least three different scenarios
are possible for the sound speed as a function of density: \textit{i)}
monotonic and sub-conformal: $c_s^2<1/3$; \textit{ii)} nonmonotonic and
sub-conformal: $c_s^2<1/3$; \textit{iii)} nonmonotonic and sub-luminal:
$c_s^2\leq1$. Clearly, scenarios \textit{ii)} and \textit{iii)} imply at
least one local maximum in the sound speed, which in \textit{iii)} can
reach values larger than $1/3$.  That the sound speed is small at low
densities and approaches the conformal value at asymptotically large
densities from below could be seen as evidence for an universal bound
$c_s^2< 1/3$, thus favouring scenario \textit{i)} and \textit{ii)}.
However, as already pointed out by~\citet{Bedaque2015}, this bound is in
tension with the most massive neutron stars observed and by a number of
counter examples in QCD at large isospin
density~\citep{Carignano:2016lxe}, two-color QCD \citep{Hands:2006ve},
quarkyonic matter~\citep{McLerran:2018hbz, Margueron:2021dtx,
  Duarte:2021tsx}, models for high-density QCD~\citep{Pal:2021qav,
  Braun:2022, Leonhardt:2019fua} and models based on the gauge/gravity
duality~\citep{Ecker:2017fyh, Demircik:2021zll, Kovensky:2021kzl}.  All
of these example favour scenario \textit{iii)}. Finally, astrophysical
measurements of neutron-star masses
$M\gtrsim2~M_\odot$~\citep{Antoniadis_fulllist:2013, Cromartie2019,
  Fonseca2021} and theoretical predictions on the maximum
(gravitational) mass~\citep{Margalit2017, Rezzolla2017, Ruiz2017,
  Shibata2019, Nathanail2021}, suggest stiff EOSs with $c_s^2\gtrsim 1/3$
at densities $\gtrsim n_0$, again selecting the scenario \textit{iii)} as
the most likely (see also~\cite{Moustakidis2017, Kanakis-Pegios:2020jnf},
who study upper limits on $c_s^2$ by extending various EOSs with a
maximally stiff parametrization at high densities).

In this \textit{Letter}, we investigate which of these scenarios is the
most natural by exploiting an agnostic approach in which we build a very
large variety of EOSs that satisfies constraints from particle theory and
astronomical observations. More specifically, we construct the EOSs using
the sound-speed parametrization introduced by~\citet{Annala2019}, which
avoids discontinuities in the sound speed as those appearing in a
piecewise polytropic parametrization~\cite{Most2018} (see also
alternatives that guarantee smoothness, like the spectral approach
of~\citet{Lindblom2012}, or the Gaussian parametrizations
by~\citet{Greif2019}). We then compute the probability density function
(PDF) of the sound speed as derived from the $1.7\times 10^6$ EOSs that
satisfy all the constraints. The behaviour of the PDF then provides a
very effective manner to determine which of the scenarios mentioned above
is the most natural given (micro)physical and (macro-)astrophysical
constraints.

\section{Methods}

The EOSs we construct are a patchwork of several different components
(see Appendix for a schematic diagram).  First, at densities $n<0.5\,n_0$
we use a tabulated version of the Baym-Pethick-Sutherland (BPS)
model~\cite{Baym71b}. Second, in the range $0.5\,n_0<n<1.1\,n_0$ we
construct monotropes of the form $p(n)=K\, n^\Gamma$, where $K$ is fixed
by matching to the BPS EOS and sample uniformly $\Gamma\in[1.77,3.23]$ to
ensure that the pressure remains entirely between the soft and stiff EOSs
of~\cite{Hebeler:2013nza}. Third, between $1.1\,n_0 < n \lesssim 40\,n_0$
we use the parametrization method introduced
by~\citet{Annala2019}, which uses the sound speed as function of the chemical
potential $\mu$ as a starting point to construct thermodynamic
quantities. In this way, the number density can be expressed as
\begin{equation}
  \label{eq:n}
 n(\mu)=n_1\,\exp \left({\int_{\mu_1}^\mu \frac{ d \mu^\prime}{\mu^\prime
     c_s^2(\mu^\prime)}}\right)\,,
\end{equation}
where $n_1=1.1\,n_0$ and $\mu_1=\mu(n_1)$ is fixed by the corresponding
chemical potential of the monotrope. Integrating Eq.~\eqref{eq:n} then
gives the pressure
\begin{equation}\label{eq:p}
 p(\mu)=p_1+\int_{\mu_1}^\mu  d \mu^\prime n(\mu^\prime)\,,
\end{equation}
where the integration constant $p_1$ is fixed by the pressure of the
monotrope at $n=n_1$. We integrate Eq.~\eqref{eq:p} numerically, using a
fixed number $N$ ($3,4,5,7$) of piecewise linear segments for the sound
speed of the following form
\begin{equation}\label{eq:cs2}
 c_s^2(\mu)=\frac{\left(\mu _{i+1}-\mu \right) c_{s,i}^2+\left(\mu -\mu
   _i\right) c_{s,i+1}^2{}}{\mu _{i+1}-\mu _i}\,,
\end{equation}
where $\mu_i$ and $c_{s,i}^2$ are parameters of the $i$-th segment in the
range $\mu_i\le \mu \le \mu_{i+1}$. The values of $c_{s,1}^2$ and $\mu_1$
are fixed by the monotrope. Note that we do not sample
$c_{s,i}^2\in[0,1]$, since this would lead to a statistical suppression
of the sub-conformal EOSs with large number of segments. Rather, for each
EOS, we first choose randomly the maximum sound speed $c_{s,{\rm
    max}}^2\in[0,1]$ and then uniformly sample the remaining free
coefficients $\mu_i\in[\mu_1,\mu_{N+1}]$ and $c_{s,i}^2\in[0,c_{s,{\rm
      max}}^2]$ in their respective domains.

As the final step in our procedure, we keep solutions whose pressure,
density and sound speed at $\mu_{N+1}=2.6\,{\rm GeV}$ are consistent with
the parametrized perturbative QCD result for cold quark matter in beta
equilibrium~\citep{Fraga2014}
\begin{equation}
  \label{eq:pQCD}
p_{_{\rm QCD}}(\mu,X) :=\frac{\mu^4}{108\pi^2}\left(c_1-\frac{d_1
  X^{-\nu_1}}{\mu/{\rm GeV}-d_2 X^{-\nu_2}}\right)\,,
\end{equation}
where $c_1=0.9008$, $d_1=0.5034$, $d_2=1.452$, $\nu_1=0.3553$,
$\nu_2=0.9101$, and the renormalization scale parameter $X\in[1,4]$. All
of the results presented in the main text refer to $1.5\times10^7$ EOSs
constructed with $N=7$ segments (see Appendix for details).

To impose the observational constraints, we compute the mass-radius
relation for each EOS by numerically solving the
Tolman-Oppenheimer-Volkoff (TOV) equations; since we construct
$\simeq10^2$ TOV solutions for each EOS, we can count on a statistics of
$\simeq1.2\times10^8$ nonrotating stellar models. We impose the mass
measurements of J0348+0432~\citep{Antoniadis_fulllist:2013} ($M=2.01\pm
0.04~M_\odot$) and of J0740+6620~\citep{Cromartie2019,
  Fonseca2021} ($M=2.08 \pm 0.07 M_\odot$) by rejecting EOSs that
have a maximum mass $M_{_{\rm TOV}}<2.0\,M_\odot$.  In addition, we
impose the radius measurements by NICER of J0740+6620~\citep{Miller2021,
  Riley2021} and of J0030+0451~\cite{Riley2019, MCMiller2019b}
by rejecting EOSs with $R<10.75\,{\rm km}$ at $M=2.0\,M_\odot$ and
$R<10.8\,{\rm km}$ at $M=1.1\,M_\odot$, respectively (see
Fig.~\ref{plot:method} in the Appendix). Finally, we exploit the
detection of GW171817 by LIGO/Virgo to set an upper bound on the binary
tidal deformability $\tilde \Lambda<720$ (low-spin
priors)~\citep{Abbott2018a}. Denoting respectively with $M_{i}$, $R_{i}$,
and $\Lambda_{i}$ the masses, radii, and tidal deformabilities of the
binary components, where
$\Lambda_i=\tfrac{2}{3}k_2\left({R_i}/{M_i}\right)^5$, $i=1,2$, and $k_2$
is the second tidal Love number, we compute the binary tidal
deformability as
\begin{equation}
\tilde{\Lambda} := \frac{16}{13} \frac{ \left( 12M_2 + M_1\right) M_{1}^4
  \Lambda_1 + \left(12M_1 + M_2 \right)M_{2}^4 \Lambda_2}{\left( M_1 +
  M_2\right)^5}\,.
\end{equation}
For any choice of $M_{1,2}$ and $R_{1,2}$, we then reject those EOSs with
$\tilde\Lambda>720$ for a chirp mass $\mathcal{M}_{\rm chirp}:=(M_1
M_2)^{3/5}(M_1+M_2)^{-1/5}=1.186 M_{\odot}$ and $q:={M_2}/{M_1}>0.73$ as
required for consistency with LIGO/Virgo data for
GW170817~\citep{Abbott2018b}.
  
\section{Results}

In order to build the various PDFs, we discretize the
corresponding two-dimensional space of solutions (\eg in the $(M,R)$
space) in $700\times700$ equally spaced cells (either linearly or
logarithmically), count the number of curves that cross a certain cell
and normalize the result by the maximum count on the whole
grid. Because the normalization is made in two dimensions, slices
  along a fixed direction do not yield normalized distributions.

\begin{figure}[htb]
\center
\includegraphics[width=0.48\textwidth]{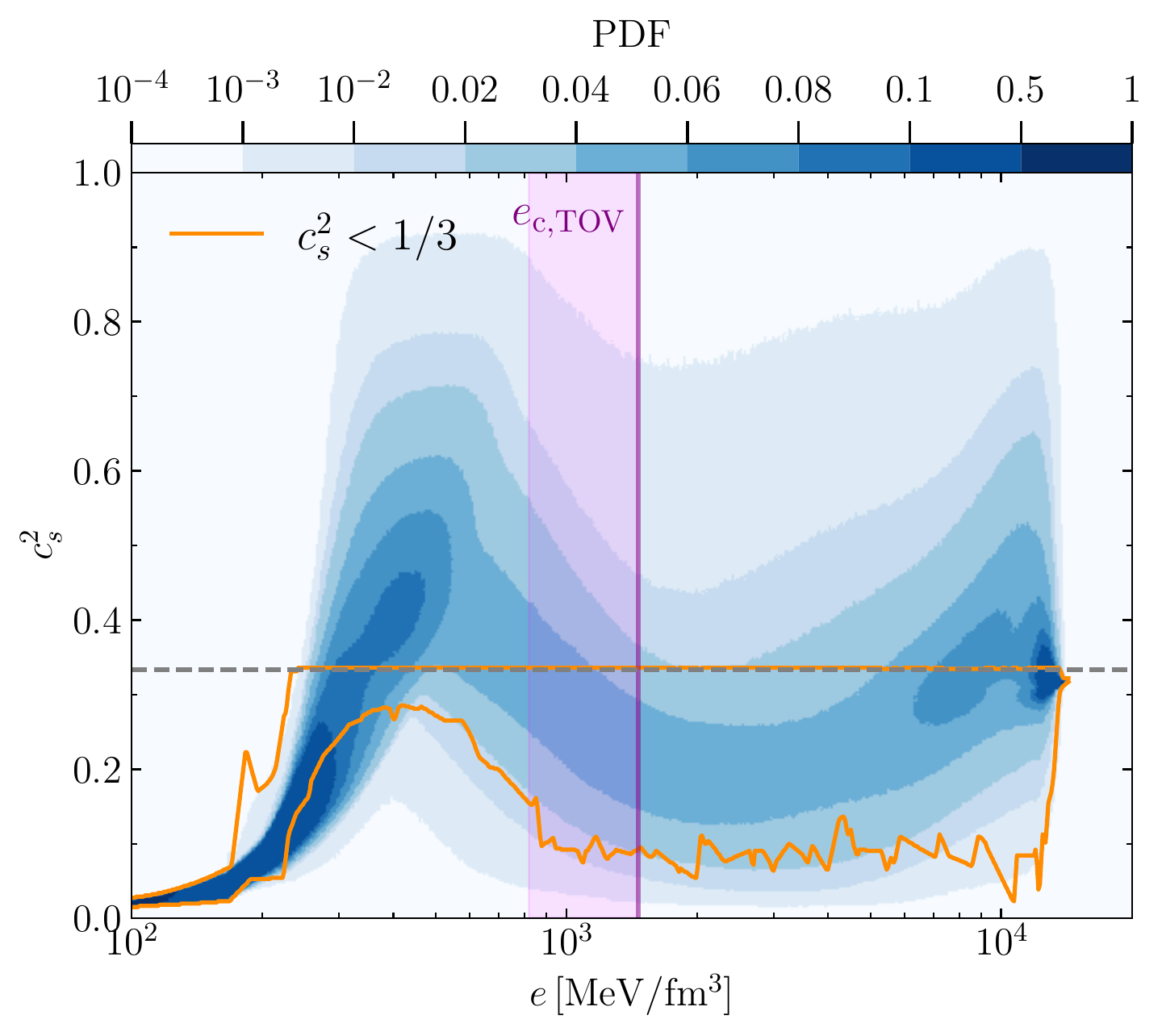}
\caption{PDF of the sound speed squared as function of the energy
  density. The purple region marks the $95\%$-interval of maximum central energy
  densities, so that the vertical purple line represents an estimate for 
  the largest possible energy density in a neutron star.
  The orange contour marks the region containing EOSs with $c^2_s<1/3$.}
\label{plot:cs2e}
\end{figure}

Figure~\ref{plot:cs2e} shows the PDF of $c^2_s$ as a function of the
energy density, with the purple region marking the $95\%$ range of
maximum central energy densities, that is, the central energy density
reached by any EOS by the star with the maximum mass $M_{_{\rm
    TOV}}$. Stated differently, the right edge of the purple region
(vertical purple line) marks our estimate for the largest possible energy
density encountered in a neutron star; in our sample, we obtain the
median $e_{c,{\rm TOV}}=1064^{+399}_{-244}\,{\rm MeV}/{\rm fm}^3$ at
$95\%$ confidence.

Note that the PDF shows a steep increase to $c_s^2\gtrsim 1/3$ for $e
\lesssim 500\,{\rm MeV}/{\rm fm}^3$, thus signalling a significant
stiffening of the EOS at these densities and a subsequent decrease of the
sound speed for larger energy densities. As a result, the PDF illustrates
how a nonmonotonic behaviour is most natural for the sound speed, hence
how the physical and observational constraints favour scenario
\textit{iii)}.  Models for quarkyonic matter~\citep[see,
  \eg][]{McLerran:2018hbz}) typically show a peak at low densities
similar to the one in our PDF~\citep{Hippert:2021}.

The orange line in Fig.~\ref{plot:cs2e} marks the region of the EOSs that
are sub-conformal, \ie with $c_s^2<1/3$, at all densities (the horizontal
dashed line that marks $c_s^2=1/3$). Note that around $500\,{\rm
  MeV}/{\rm fm}^3$, the orange contour spans a very thin region,
indicating that at these energy densities the sub-conformal EOSs have an
obvious upper bound $c_s^2 < 1/3$, but also a less-obvious lower bound
$c_s^2\gtrsim 0.2$. This is an important feature that explains why these
EOSs are so difficult to produce. Indeed, as revealed by the colormap,
the number of EOSs that fall in this region is very small and amounts to
only $\simeq 5\times10^{-5}$ of the total. The fraction of sub-conformal
EOSs increases slightly if we restrict the range of densities to those
that are admissible for neutron-star interiors, becoming $\simeq
3\times10^{-4}$ of the total. The reason for this increase is that many
of the EOSs that are sub-conformal for $e \lesssim 300\,{\rm
  MeV/fm^{-3}}$, tend to stiffen at larger energy densities, thus
becoming super-conformal.

The colormap of the PDF in Fig.~\ref{plot:cs2e} also reveals the presence
of a second peak at large energy densities, close to where the
perturbative QCD boundary conditions are imposed and reflects artefacts
of the parametrization method, which allows for large variations in the
sound speed at very high energy densities, where $c_s^2$ is expected to
be close to the asymptotic value $1/3$. Fortunately, the energy densities
where this second peak appears are far from those expected in the
interior of neutrons stars. It is also possible to reduce the extent of
this second peak by imposing a criterion that filters out EOSs whose
sound speeds vary strongly on small scales as done
by~\citet{Annala2019}. However, given the very poor knowledge of the
behaviour of the sound speed at these regimes, we prefer to report the
unfiltered results. What matters here is that, when imposed, the
filtering has no significant impact on the PDF at the energy densities
that are relevant for stellar interiors (see the Appendix for a discussion).

\begin{figure}
\center
\includegraphics[width=0.48\textwidth]{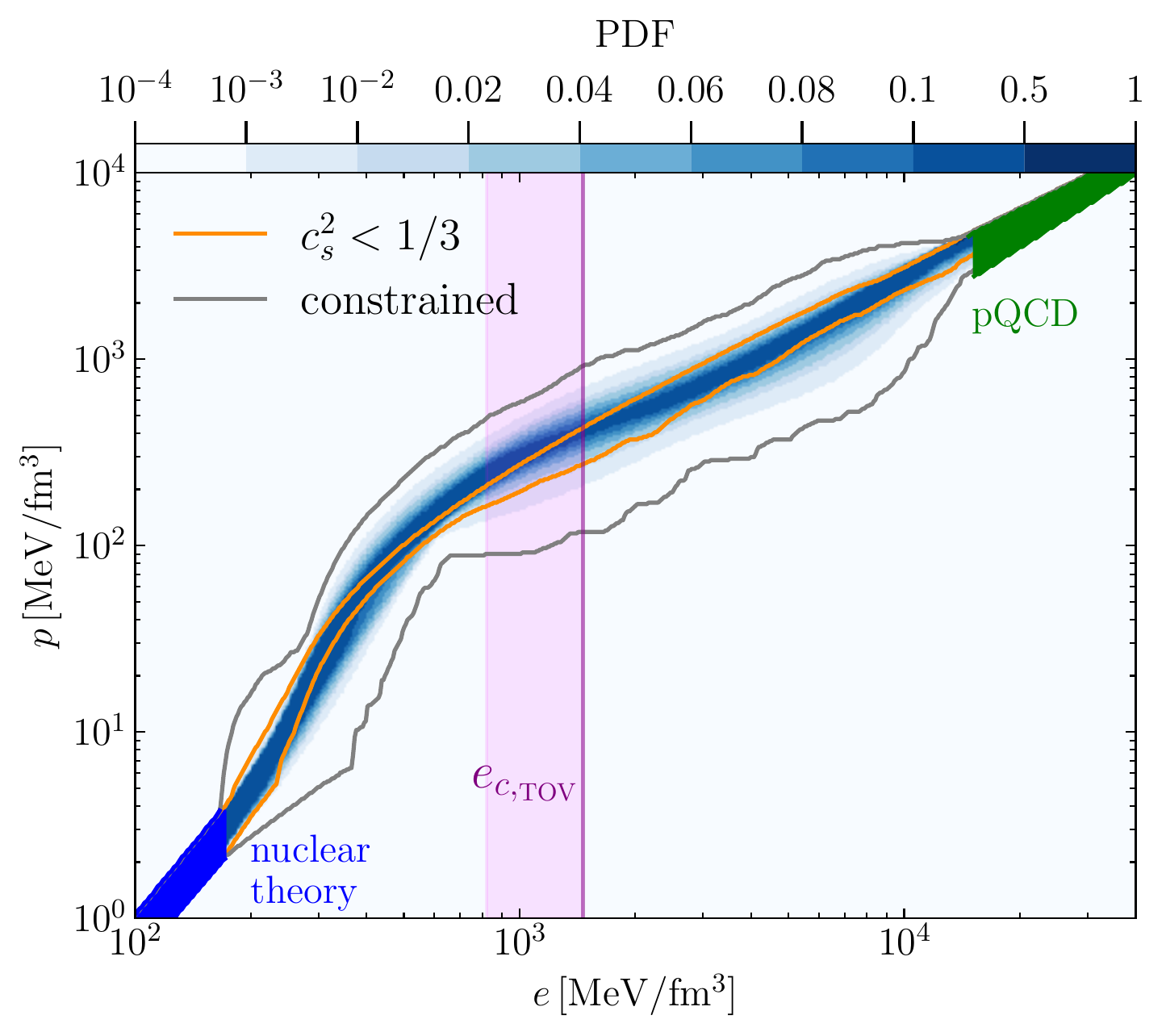}
\caption{PDF of the pressure as function of the energy density. The blue
  and green regions mark the uncertainties in pressure in nuclear
  theory~\citep{Hebeler:2013nza} and perturbative QCD~\citep{Fraga2014},
  respectively. The gray contour encloses all EOSs satisfying the
  astronomical constraints, while the purple regions and the orange
  contour are the same as in Fig.~\ref{plot:cs2e}.}
\label{plot:EOS}
\end{figure}

Figure~\ref{plot:EOS} shows the corresponding PDF of the pressure as a
function of the energy density with the same conventions as in
Fig.~\ref{plot:cs2e}. In addition, we indicate with a gray line the outer
envelope of all constraint satisfying solutions, which is very similar to
the one found by~\citet{Annala2019}. However, an important difference
with respect to~\citet{Annala2019}, where no information on the
distribution is offered, is that the PDF reveals that the large majority
of EOSs accumulates in a band that is much narrower than the allowed gray
region. Interestingly, at almost all energy densities, the pressure is
either inside or close to the sub-conformal region, except around
$e\approx 500\,{\rm MeV}/{\rm fm}^3$.  This is an important result as it
reveals that it is very hard to deduce the poor likelihood of
sub-conformal EOSs when looking at the behaviour of the pressure only.

\begin{figure}
\center
\includegraphics[width=0.48\textwidth]{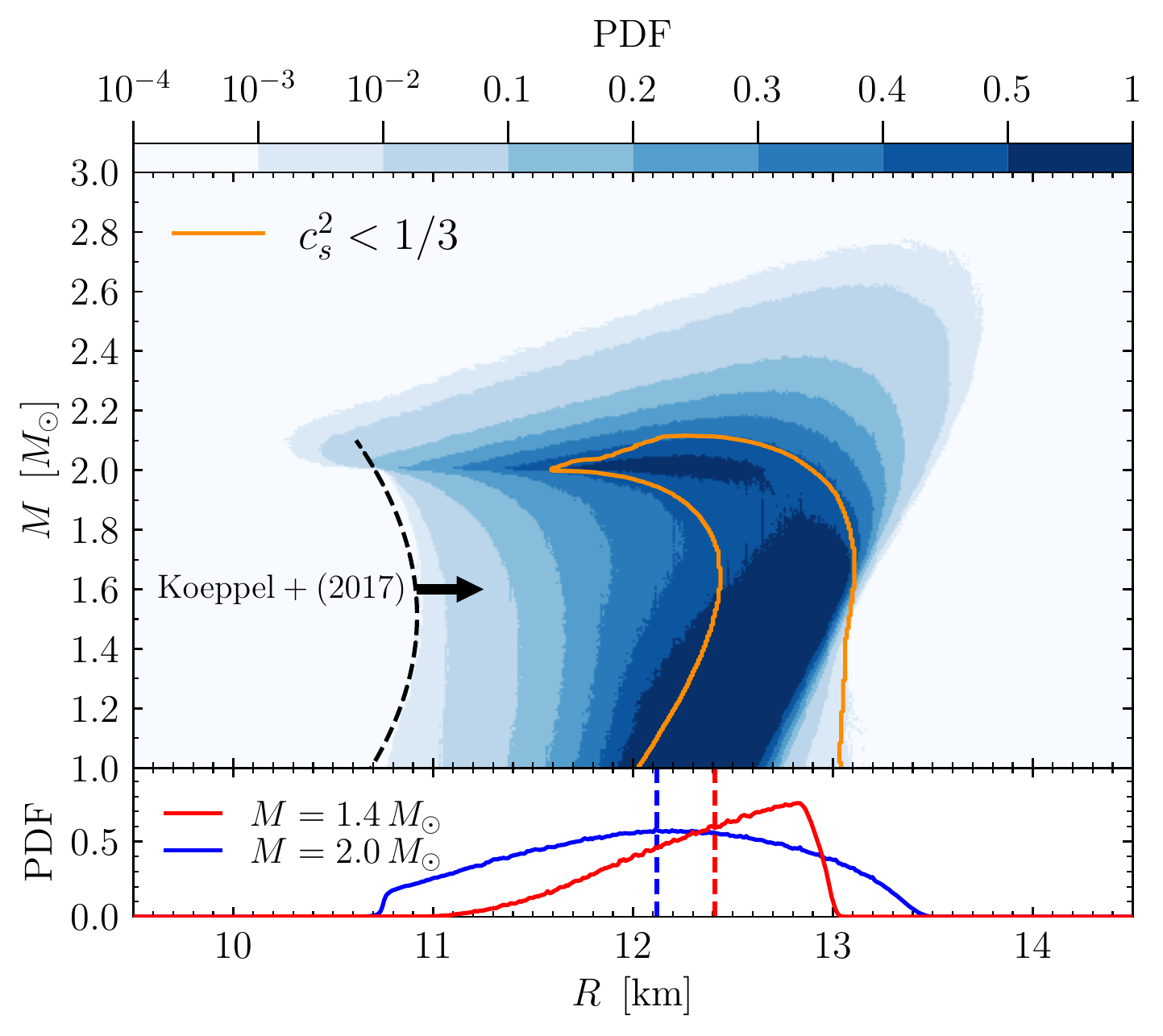}
\caption{Top: PDF of the mass-radius relation.  The orange contour
  encloses EOSs with sub-conformal sound speed (see
  Fig.~\ref{plot:cs2e}). The dashed black line reports the lower bound
  for the radii as computed from considerations on the threshold
  mass~\citep{Koeppel2019}. Bottom: PDF cuts for stars with mass
  $M=1.4\,M_\odot$ and $M=2\,M_\odot$; the dashed vertical lines mark the
  median of the distributions.}
\label{plot:MRd}
\end{figure}

Figure~\ref{plot:MRd} shows the PDF of the masses as a function of the
stellar radii. The outer edges of the distribution show $M_{_{\rm
    TOV}}\lesssim 3\,M_\odot$ and an approximate lower limit for the
minimum radius of $R \gtrsim 10.5\,{\rm km}$. This is in remarkably good
agreement with the analytic lower bound $R/{\rm km}\gtrsim
-0.88\,(M/M_{\odot})^2 +2.66\,(M/M_{\odot})+8.91$ derived when using the
detection of GW170817 and the estimates on the threshold mass to prompt
collapse~\citep{Koeppel2019, Tootle2021}. The orange line marks again the
outer envelope spanned by sub-conformal EOSs, whose maximum mass is
$M\approx 2.1\,M_\odot$, also seen by~\citet{Annala2019, 
  Annala:2022}. This behaviour confirms on rather general grounds the
tension between $c_{s,\rm max}^2<1/3$ and the observation of stars with
$M\gtrsim 2~M_\odot$, already been pointed out~\citep[see,
  \eg][]{Bedaque2015, Alford2015,Tews:2018kmu}.

In the bottom panel of Fig.~\ref{plot:MRd} we report slices of the PDF
for two selected masses, namely, $1.4$ and $2.0\,M_{\odot}$. The median
values (dashed lines) and the $95\%$ confidence levels allow us to
estimate the corresponding radii as $R_{1.4}=12.42^{+0.52}_{-0.99}\,{\rm
  km}$ and $R_{2.0}=12.11^{+1.11}_{-1.23}\,{\rm km}$, respectively. The
significant skew in the distribution for $M=1.4\,M_\odot$ is due to the
tidal-deformability constraint which penalizes large radii. Despite the
different method, our estimate for the median of $R_{1.4}$ is in good
agreement with the piecewise polytropic estimates of~\citet{Most2017}
($12.00 < R_{1.4} < 13.45\,{\rm km}$) and slightly larger, but well
within the error bars of the estimates by~\citet{Huth2022}~\citep[see
  also][for similar estimates]{Abbott2018b, Radice2018c,
  Capano2020, Dietrich2020, Biswas2021}.

Interestingly, we find that none of our constraint-satisfying EOSs have
monotonic sound speeds, as it is required by scenario \textit{i)}. In
order to check if the scenario \textit{i)} is just unlikely or indeed
inconsistent with the constraints, we construct $1.5\times10^7$ EOSs with
monotonic sound speed, hence sampling the sound speed in $c^2_s \in [0,
  1/3]$, without imposing a lower bound on $M_{_{\rm TOV}}$. We then find
that these EOSs are only able to support a maximal mass $M_{_{\rm
    TOV}}<1.99\,M_\odot$ and are therefore excluded by a two-solar mass
bound (see the Appendix for the PDF of this constraint-violating subset).

\begin{figure}
\center
\includegraphics[width=0.48\textwidth]{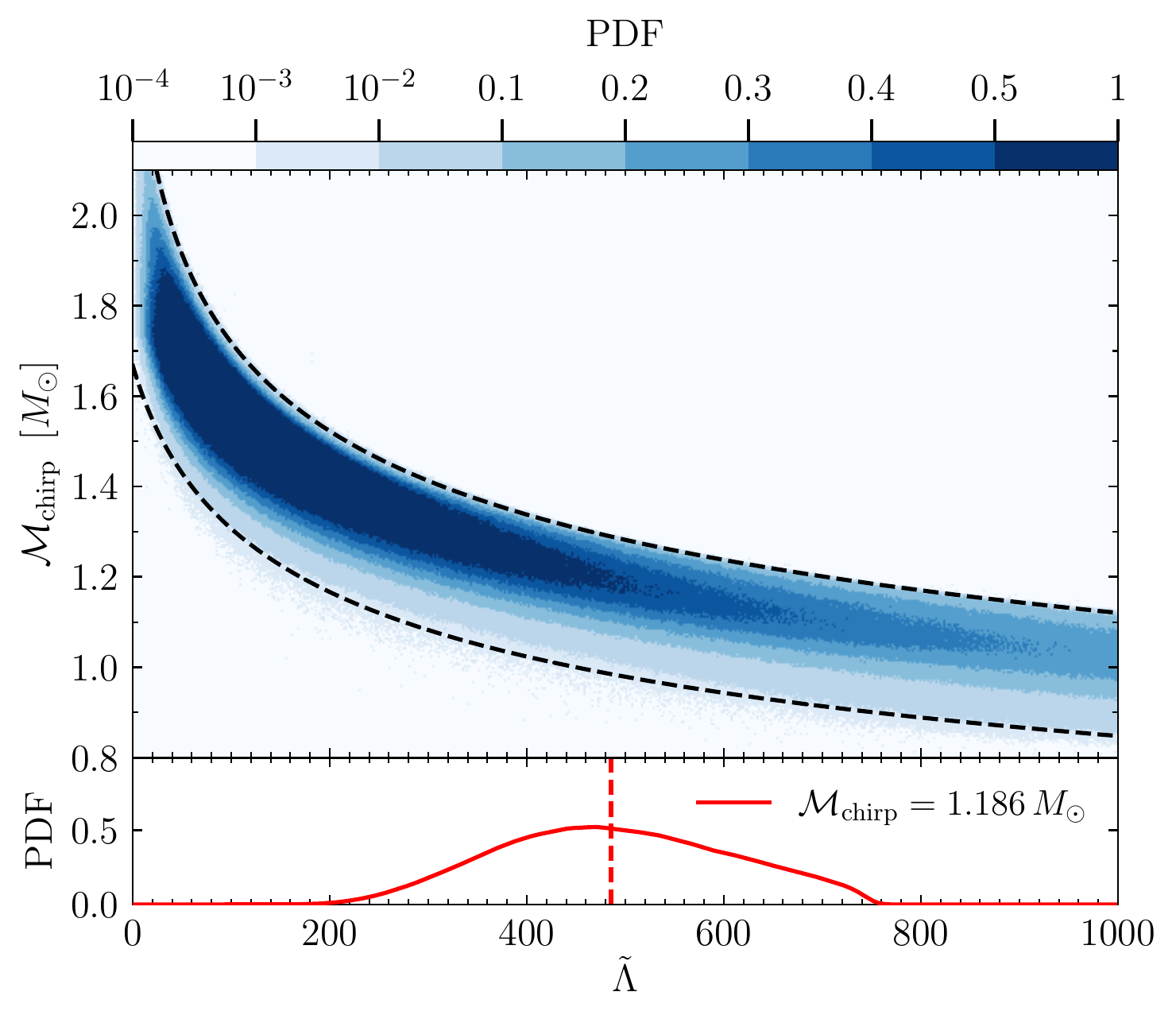}
\caption{Top: PDF of the chirp mass as a function of the binary tidal
  deformability. Shown as black dashed lines are analytic fits for the
  minimum and maximum values of $\tilde{\Lambda}$ for a given chirp
  mass. Bottom: PDF cut for a binary with the same chirp mass as in
  GW170817, $\mathcal{M}_{\rm chirp}=1.186\,M_\odot$; the red dashed
  vertical line marks the median.}
\label{plot:Lambda}
\end{figure}

We conclude our discussion on the statistical properties of our EOSs with
what arguably is one of the most important results of this work. More
specifically, we report in Fig.~\ref{plot:Lambda} the PDF of the binary
deformability shown as a function of the chirp mass. In essence, for any
EOS in our sample, we randomly collect $100$ pairs of
$M_{1,2},\,\Lambda_{1,2}$, out of which we compute $\mathcal{M}_{\rm
  chirp}$ and ${\tilde \Lambda}$. The results are then binned and
color-coded as in the other PDFs. We find this is a particularly
important figure as it relates a quantity that is directly measurable
from gravitational-wave observations, $\mathcal{M}_{\rm chirp}$, with a
quantity that contains precious information on the microphysics ${\tilde
  \Lambda}$. In this way, we find that the binary tidal deformability of
a GW170817-like event is constrained to be $\tilde\Lambda_{1.186} =
483^{+224}_{-210}$, at $95\%$ confidence level. The lower bound is a
prediction while the upper bound effectively reflects the constraint
imposed on $\tilde{\Lambda}$. Furthermore, we have fitted the $99\%$
confidence contours of the PDF with simple power-laws (dashed black
lines) to obtain analytic estimates for the minimum (maximum) value of
$\tilde{\Lambda}$ as function of the chirp mass
\begin{equation}
  \label{eq:lambda_minmax}
  \tilde{\Lambda}_{\rm min\,(max)} = a + b\,\mathcal{M}_{\rm chirp}^{c}\,,
\end{equation}
where $a=-50\,(-20)$, $b=500\,(1800)$, and $c=-4.5\,(-5.0)$. The
relevance of expressions \eqref{eq:lambda_minmax} is that, once a merging
neutron-star binary is detected and its chirp mass is measured
accurately, it is straightforward to use Eq.~\eqref{eq:lambda_minmax} to
readily obtain a lower and an upper bound on the binary tidal
deformability of the system. This has not been possible so far, where
only upper limits have been set~\citep{Abbott2018b}.

\begin{figure}
\center
\vspace{5mm}
\includegraphics[width=0.45\textwidth]{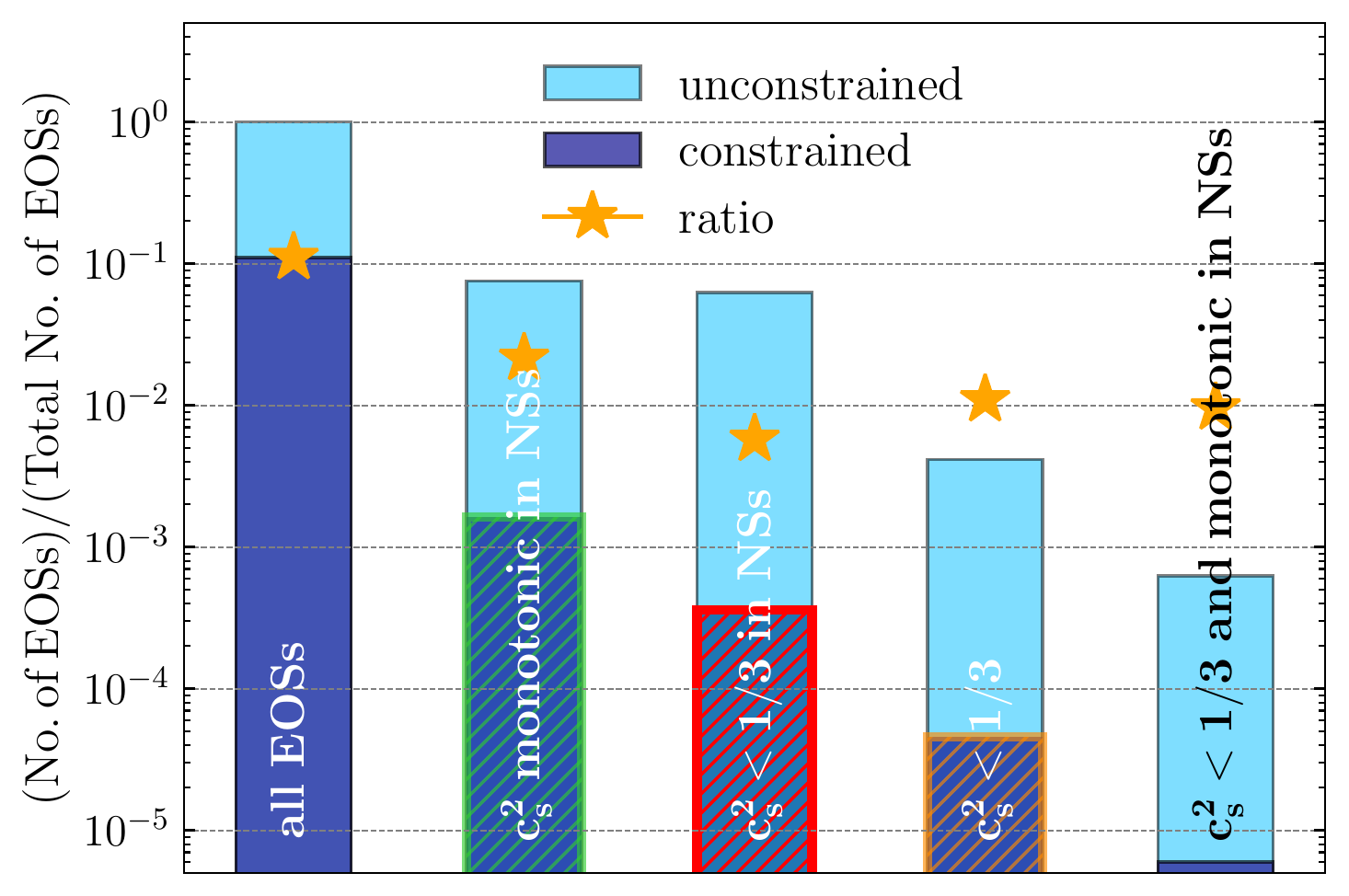}
\caption{Schematic representation of the various sets into which our
  total sample of EOSs can be decomposed, either when subject to the
  observational constraints (dark blue) or when not (light
  blue). The ratio between the two samples is shown with a
    stellar marker, while the hatched color-coding reflects that used in
  Fig.~\ref{plot:cs2e} and in the Appendix.}
\label{plot:count}
\end{figure}

Finally, to provide a graphical representation of our statistics, we show
in Fig.~\ref{plot:count} the relative weight of the various sets into
which our total sample of EOSs can be decomposed, either when subject to
the observational constraints (dark blue) or when not (light blue). Note
that the heights of the light-blue bars in Fig.~\ref{plot:count} for
scenarios \textit{i)} (fourth bar) and \textit{iii)} (first bar) are not
the same. In other words, the unconstrained samples of these scenarios
are not equally populated. This is simply because we sample the sound
speed uniformly, so that the EOSs in which all of the values of $c^2_s$
have to fall below $1/3$, as required for scenario \textit{i)}, are
statistically suppressed when compared to the generic scenario
\textit{iii)}, where $c^2_s$ can assume values in the full range $[0,
  1]$. What matters, however, is the relative difference in the
constrained/unconstrained samples (see stellar markers), which is an
order of magnitude smaller for scenario \textit{i)} than for scenario
\textit{iii)}.  This means that \textit{iii)} is statistically preferred
over \textit{i)} by the observational data. Furthermore, we note that in
our full set we do not find any EOS with monotonic $c_s^2$ at all
densities that satisfies the observational constraints.

\section{Conclusion}

We have constructed a large number of viable EOSs and studied the
statistical distributions of the corresponding sound speed, mass-radius
relation, and binary tidal deformability. We found that a steep rise in
the sound speed and a peak $c^2_s>1/3$ is statistically preferred and
that EOSs with sub-conformal sound speeds, \ie $c^2_s \leq 1/3$, are
possible within the stars, but also that they are unlikely, although in
principle possible, being only $0.03\%$ of our sample. Hence, it is
natural to expect that $c^2_s > 1/3$ somewhere inside the neutron
stars. Furthermore, imposing sub-conformality already at the level of the
sampling has allowed us to resolve the range of these EOSs more
accurately and identify a lower bound for the sound speed
$c_s^2\gtrsim0.2$ around $e\approx 590\,{\rm MeV}/{\rm
  fm}^3$. Interestingly, we were not able to find solutions with
monotonic sound speed. Exploiting the statistical robustness of our large
sample, we have obtained estimates at $95\%$ credibility of neutron-star
radii for representative stars with masses of $1.4$ and $2.0$ solar
masses, namely, $R_{1.4}=12.42^{+0.52}_{-0.99}\,{\rm km}$,
$R_{2.0}=12.12^{+1.11}_{-1.23}\,{\rm km}$, and for the binary tidal
deformability of the GW170817, $\tilde\Lambda_{1.186}=485^{+225}_{-211}$.

Remarkably, the very agnostic predictions on the mass-radius relations
emerging from our statistics matches well the analytic predictions on the
minimum stellar radius coming from very different arguments, namely the
threshold mass to gravitational collapse. New radius and mass
measurements, as well as gravitational-wave detections of binary
neutron-star mergers will allow us to improve these estimates and narrow
down our PDFs further. For example, the discovery of a neutron-star with
mass larger than $2.1\,M_\odot$ would eliminate the possibility of
sub-conformal EOSs entirely. Future work will consider improved
constraints at neutron-star densities from perturbative QCD such as those
recently suggested by~\citet{Komoltsev:2021jzg}. In addition, alternative
approaches to the construction of the EOSs (\eg piecewise
polytropes~\citep{Most2018} or spectral
paramaterization~\citep{Lindblom2012}) will help determine a potential
bias introduced by our method, which is however expected to be much
smaller than the observational uncertainties.

\textit{Note:} After the completion of this work, a new mass measurement
of $2.35\pm0.17\,M_\odot$ has been published for the pulsar in the binary
system PSR J0952-0607 \citep{Romani:2022jhd}. 
This new mass bound has no impact on the main results and conclusions presented here. 
While a detailed discussion is beyond the scope of this paper, 
the impact of very large maximum masses on 
the internal properties of neutron stars is instead analyzed by~\citep{Ecker:2022dlg}.

\begin{acknowledgments}

We thank R.~Duque, T.~Gorda, J.~Jiang and A.~Vuorinen for valuable 
discussions and comments on the manuscript as well
as A.~Cruz-Osorio and K.~Topolski for help with the
visualizations. Partial funding comes from the State of Hesse within the
Research Cluster ELEMENTS (Project ID 500/10.006), by the ERC Advanced
Grant ``JETSET: Launching, propagation and emission of relativistic jets
from binary mergers and across mass scales'' (Grant
No. 884631). C.~E. acknowledges support by the Deutsche
Forschungsgemeinschaft (DFG, German Research Foundation) through the
CRC-TR 211 ``Strong-interaction matter under extreme conditions''--
project number 315477589 -- TRR 211.
\end{acknowledgments}

\appendix
\section{Details about EOS construction}

The material presented here is meant to provide additional information on
some of the details of our calculations and on their robustness. 

\begin{figure*}
  \includegraphics[width=0.25\textheight]{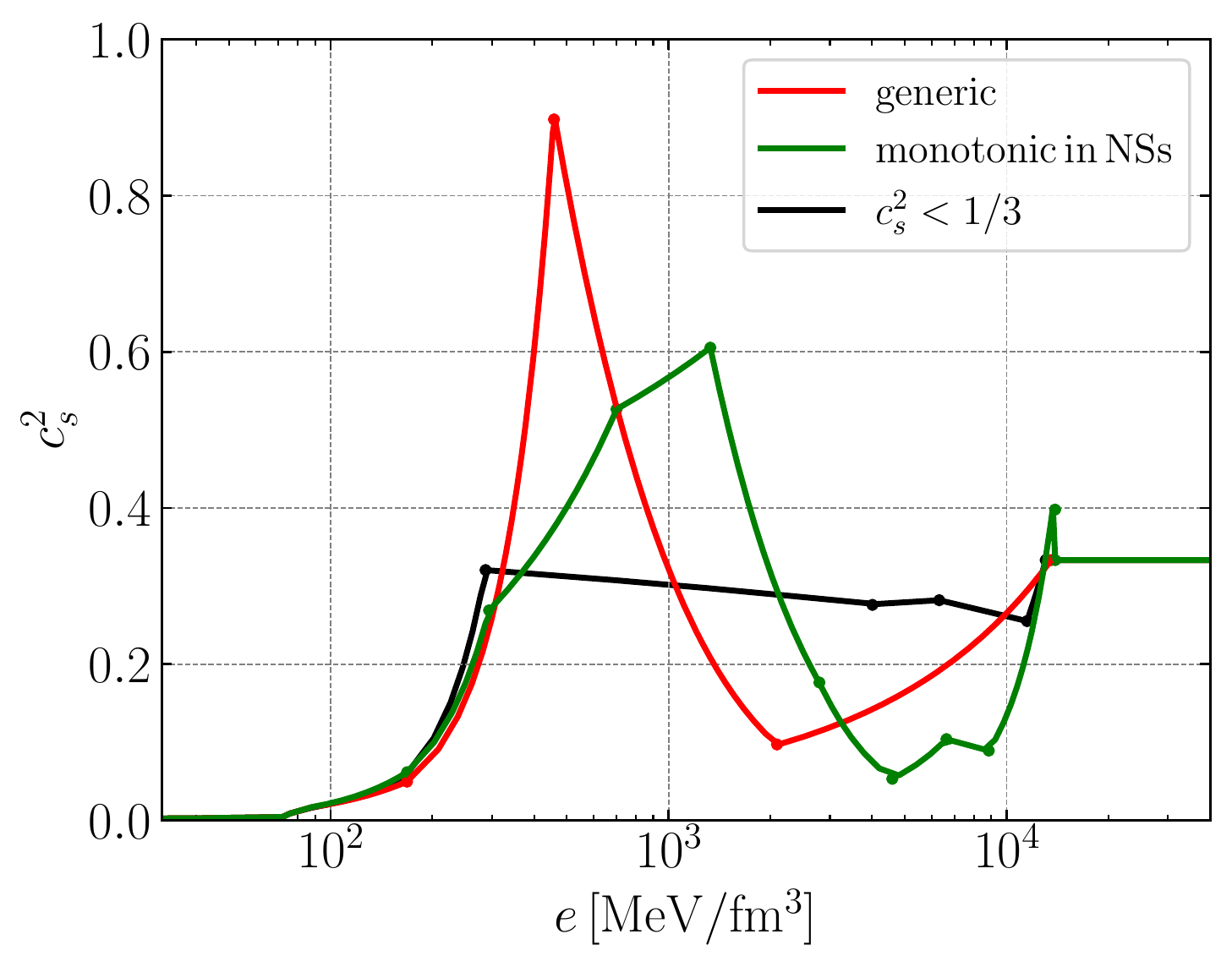}
  \includegraphics[width=0.25\textheight]{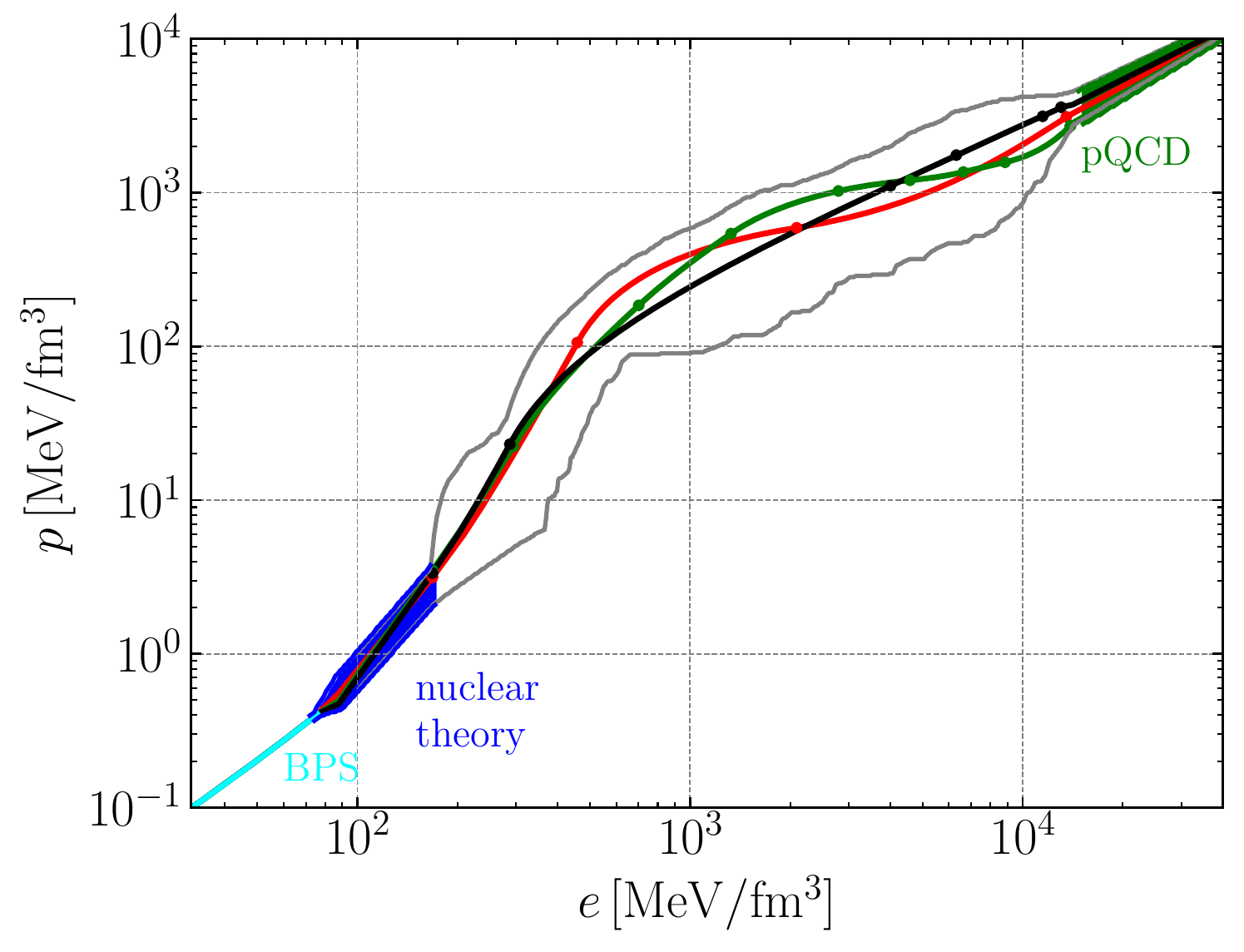}
  \includegraphics[width=0.25\textheight]{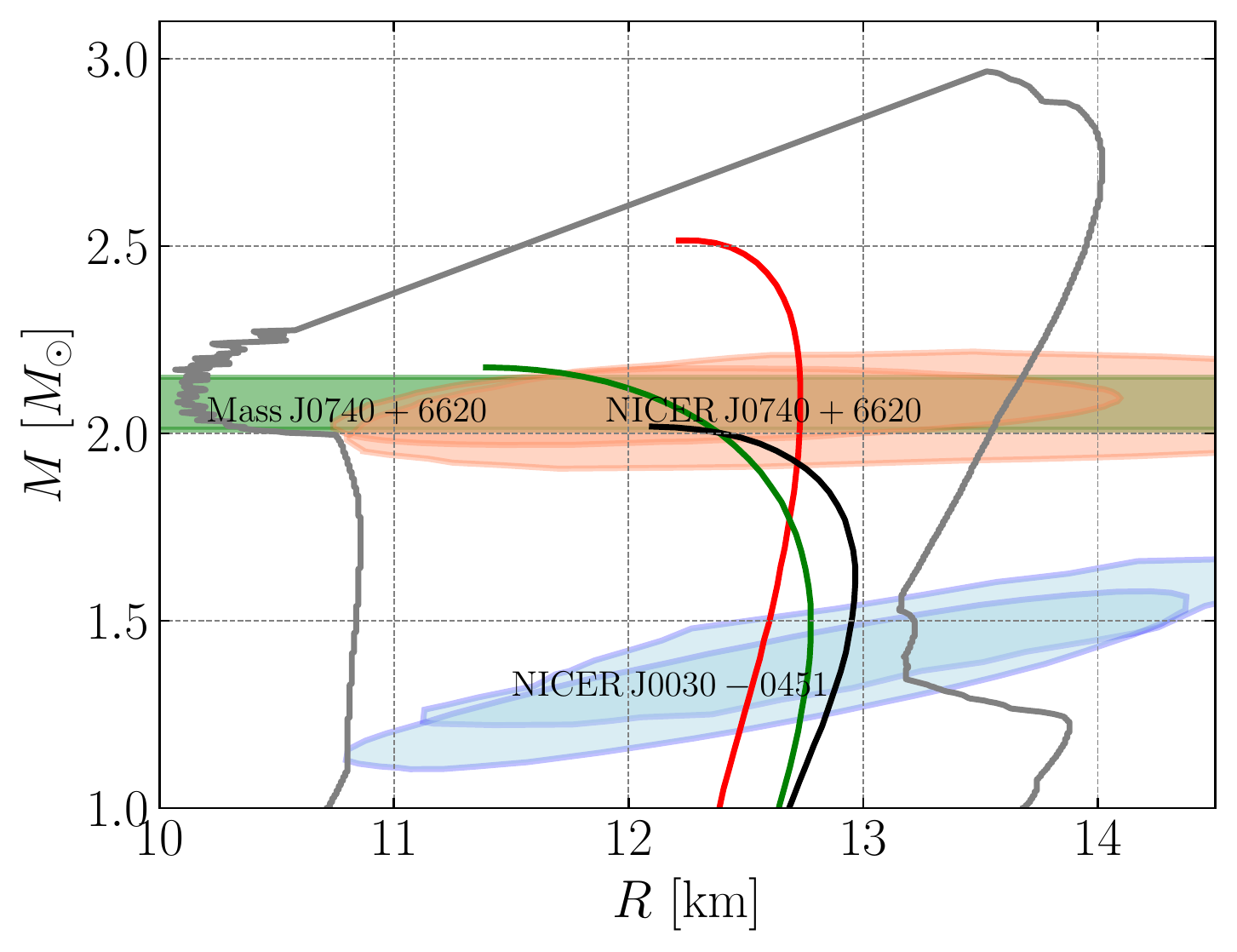}
  \caption{Representative examples referring to a generic EOS (red solid
    line), a sub-conformal EOS (blue solid line), and monotonic sound
    speed inside neutron stars (green solid line). The left plot shows
    the sound speed as function of the energy density, with the kinks in
    the curves marking the matching points of the various segment used to
    construct the EOSs. The middle panel shows the corresponding EOSs,
    together with the BPS EOS~\citep{Baym71b} (cyan solid line) that we
    use at low densities.  The blue and green bands are the uncertainties
    from nuclear theory~\citep{Hebeler:2013nza} and perturbative
    QCD~\citep{Fraga2014}, respectively. Shown in gray is the outer
    envelope of all constraint-satisfying EOSs. The right panel reports
    the corresponding mass-radius curves, together with the outer
    envelope and various observational constraints.}
\label{plot:method}
\end{figure*}

We start by showing in Fig.~\ref{plot:method} some illustrative examples 
of our method for the construction of the EOSs.
For simplicity, we concentrate on three representative examples referring to
a generic EOS (\ie an EOS with no specific behaviour of the sound speed;
red solid line), a sub-conformal EOS (\ie an EOS with $c^2_s < 1/3$; blue
solid line), and monotonic sound speed inside neutron stars (\ie an EOS
with monotonic $c^2_s$ for $e \lesssim e_{c,{\rm TOV}}\approx1.3\,{\rm
GeV}\,{\rm fm}^{-3}$ in this case; green solid line).
More specifically, the plot on the left shows the sound speed as function
of the energy density; note that the curves are continuous but with
discontinuous derivatives and that the kinks in the curves mark the
matching points of the various segment used to construct the EOSs. The
middle panel shows the corresponding EOSs, together with the BPS
EOS~\citep{Baym71b} (cyan solid line) that we use at low densities.  The
blue and green bands are the uncertainties from nuclear
theory~\citep{Hebeler:2013nza} and perturbative QCD~\citep{Fraga2014},
respectively.  We also show in gray the outer envelope of all
constraint-satisfying EOSs. Finally, the right panel shows the
corresponding mass-radius curves, together with the outer envelope and
various observational constraints~\citep{Miller2021, Riley2021,
  Antoniadis_fulllist:2013, Cromartie2019, Fonseca2021} we impose.  Note
that the gray contour reproduces the hump at large radii around $M\approx
1.2\,M_\odot$, also seen by~\citet{Annala2019}, and which was associated
to dynamically unstable solutions by~\citet{Jimenez:2021wil}.  Comparing
this part of the contour to Fig.~\ref{plot:MRd} in the main text clearly
shows that the solutions in this region are possible but very unlikely.

\begin{figure}[h!]
\center
\includegraphics[width=0.42\textwidth]{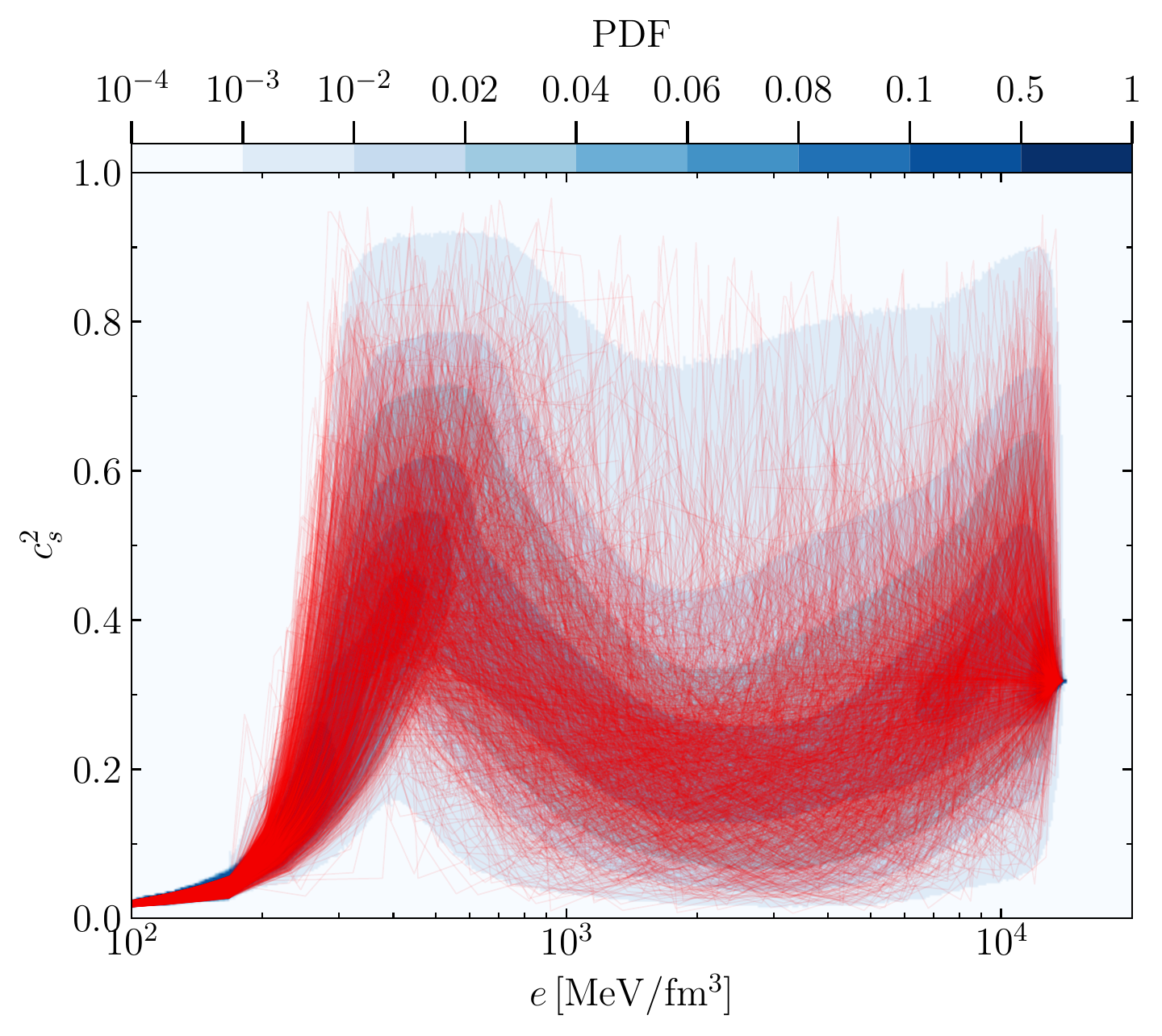}
\caption{The same as Fig. 1 in the main text, but with the overlay of the
sound-speed curves relative to 1000 different EOSs.}
\label{plot:cs2PDFcurves}
\end{figure}

\begin{figure*}
\center
\includegraphics[width=0.24\textwidth]{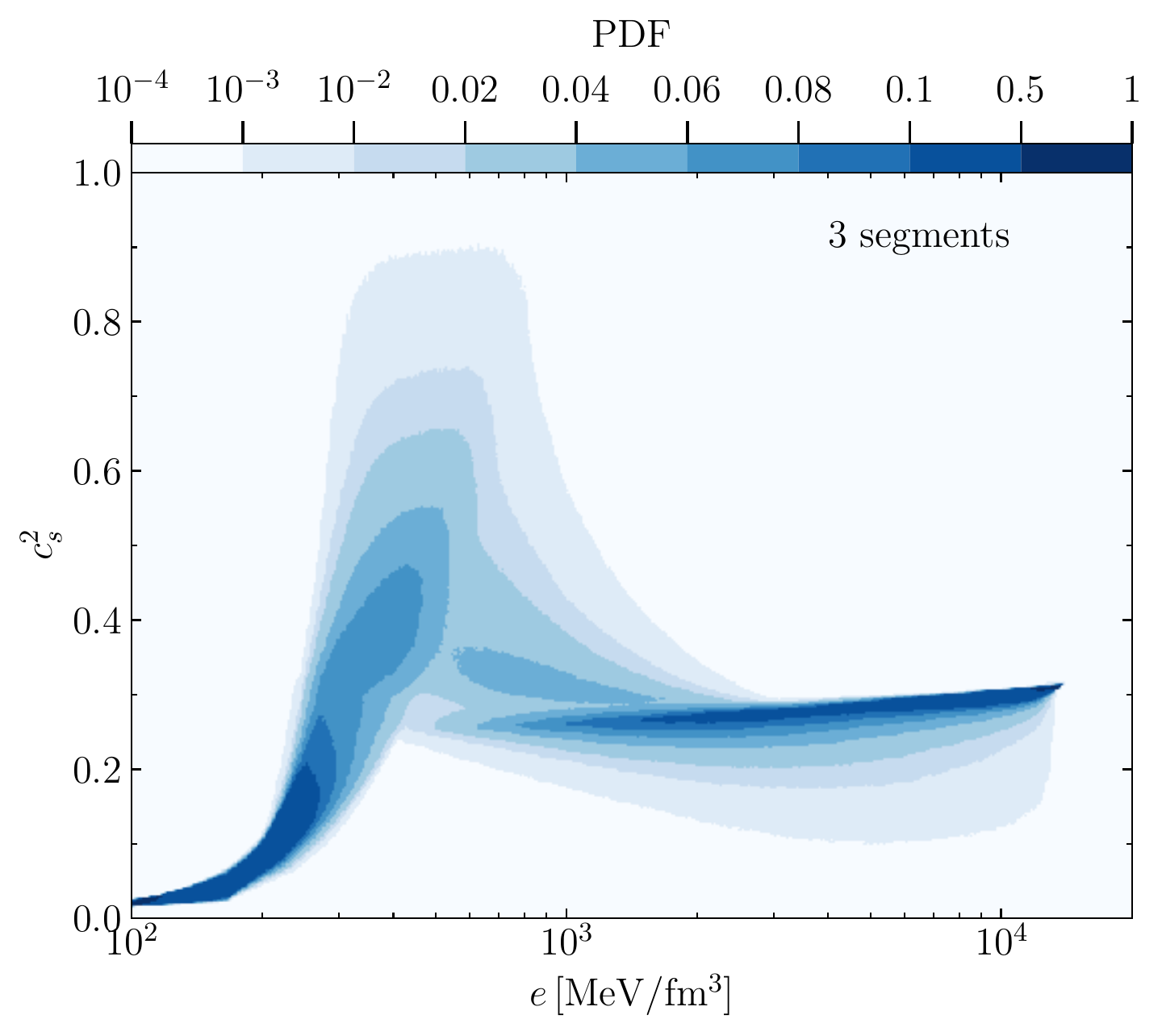}
\hskip 0.1cm
\includegraphics[width=0.24\textwidth]{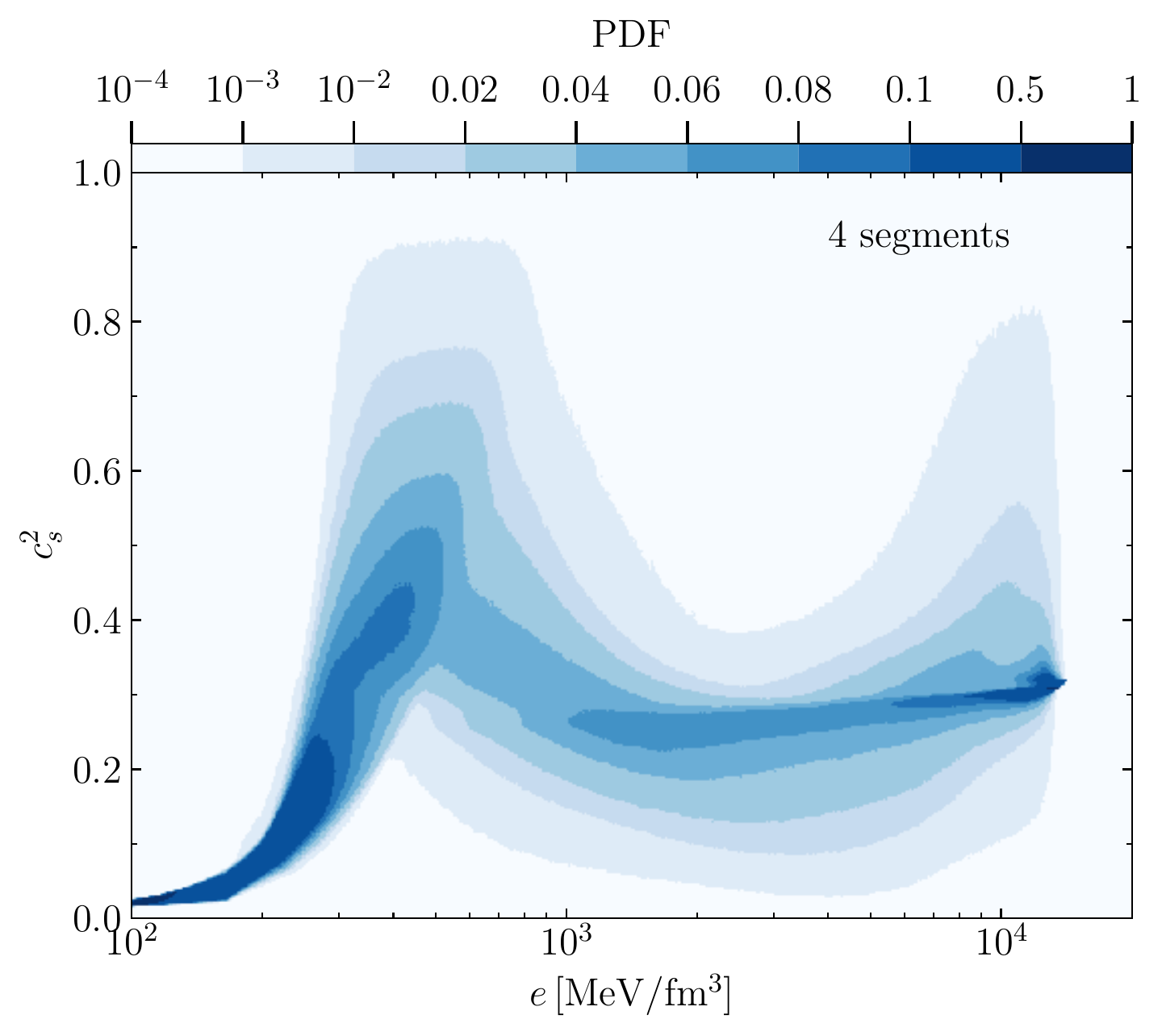}
\hskip 0.1cm
\includegraphics[width=0.24\textwidth]{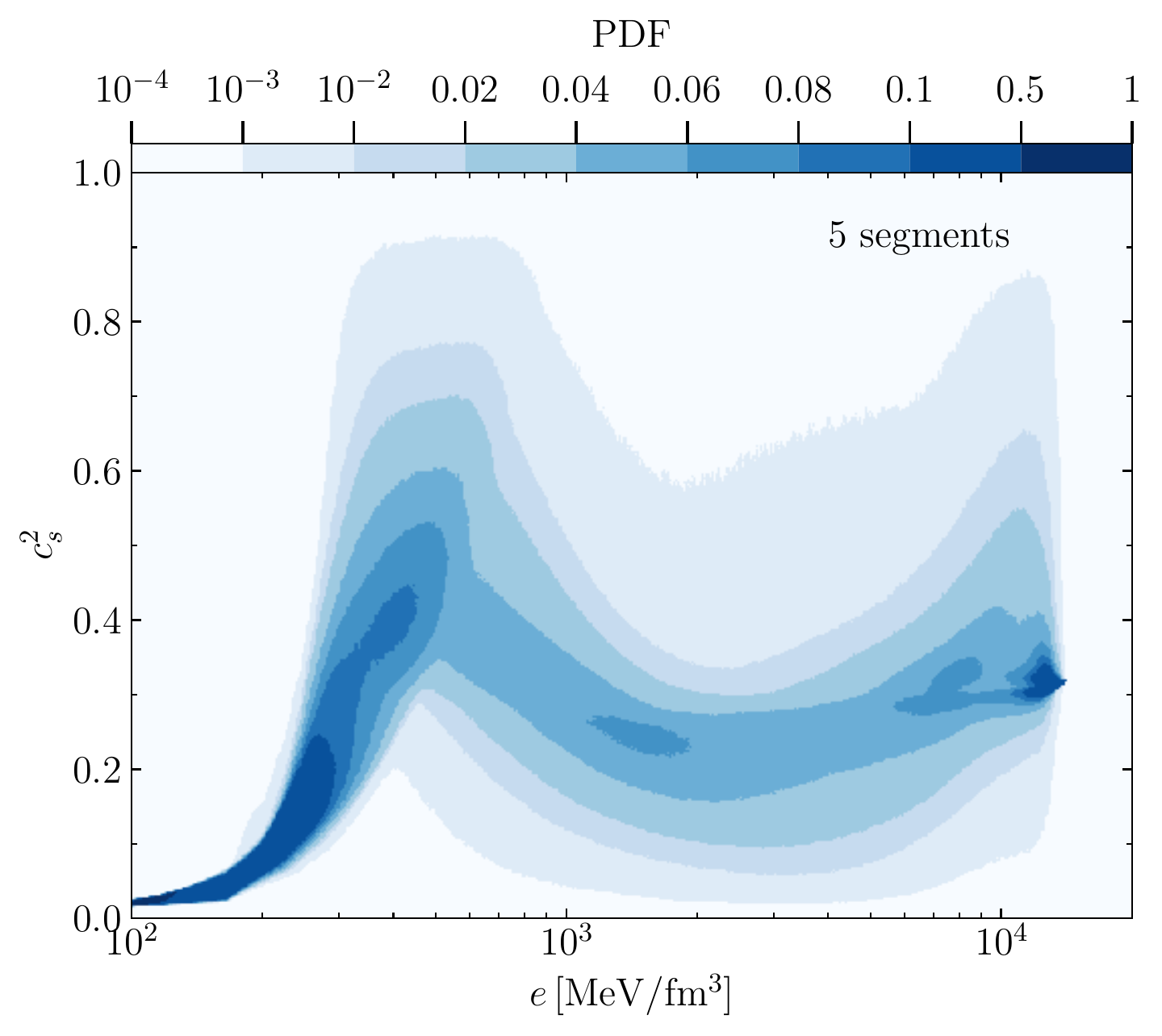}
\hskip 0.1cm
\includegraphics[width=0.24\textwidth]{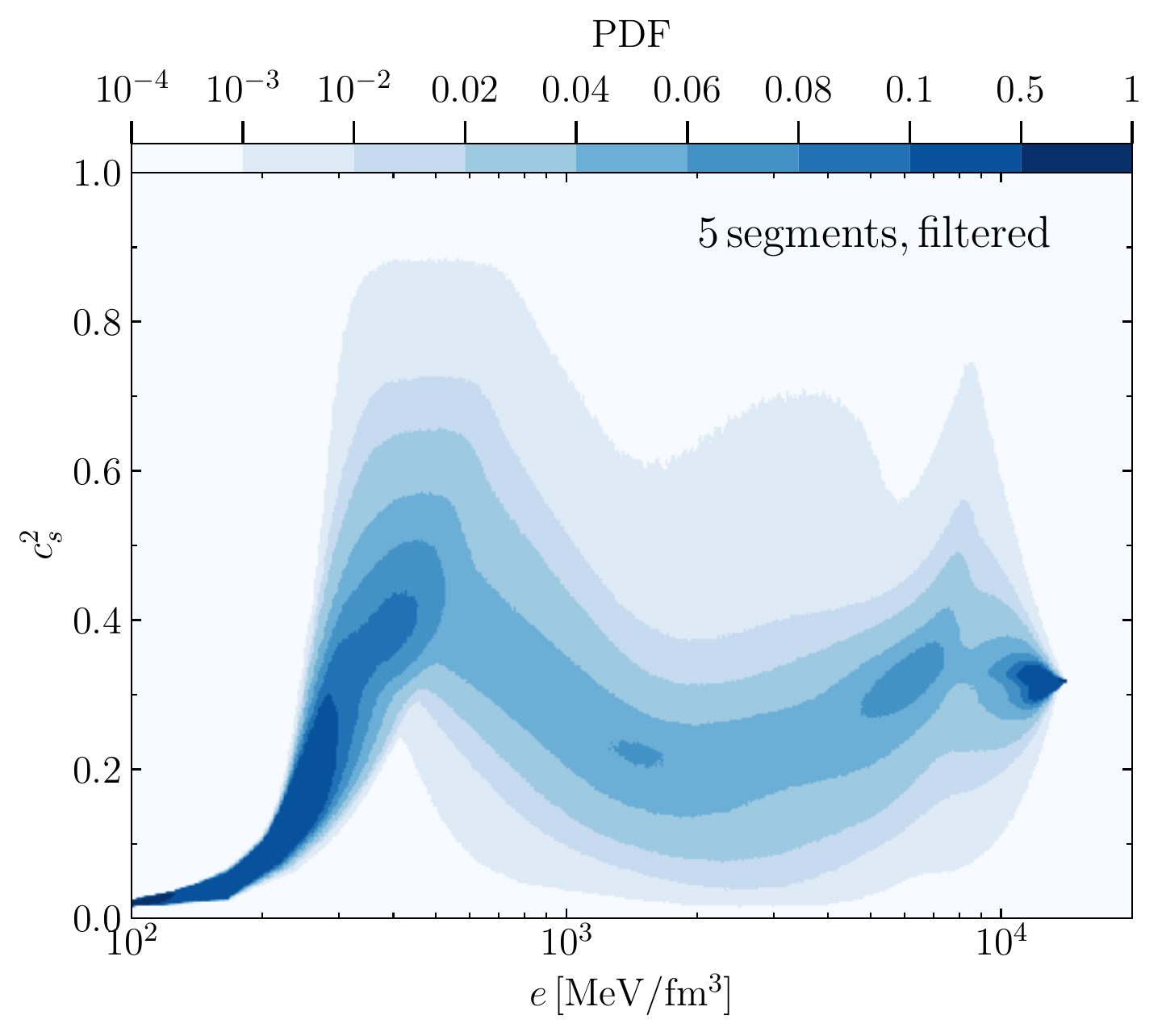}
\caption{PDFs of the sound speed obtained for different values of the
number of segments. More specifically, the first three panels from the
left refer to $N=3, 4, 5$, respectively. They clearly show that the steep
increase of the distribution beyond $c^2_s \simeq 1/3$ at small energy
densities is a robust result that is independent of the number of
segments and filtering. The right-most panel refers instead to a PDF for
$N=5$ in which the filtering on the energy-density gradient is
applied. Note how the filtering does modify the behaviour of the PDF at
very large energy densities removing the most extreme EOSs.}
\label{plot:cs2seg}
\end{figure*}

Figure~1 of the main text shows the PDF of $c^2_s$ as a function of the
energy density and a discretized colormap. Note that the steep increase
to $c_s^2\gtrsim 1/3$ for $e \lesssim 500\,{\rm MeV}/{\rm fm}^3$, thus
signalling a significant stiffening of the EOS at these densities and a
subsequent decrease of the sound speed for larger energy densities.  As a
visual aid on how this steep increase is produced,
Fig.~\ref{plot:cs2PDFcurves} is the same as Fig.~1 in the main text, but
with the overlay of the sound-speed curves relative to 1000 different
EOSs.

We next consider the impact that the number of segments employed in the
construction of the piece-wise linear segments for the calculation of the
sound speed.  As pointed out by~\citet{Annala2019}, constructing EOSs
from piece-wise linear segments for the sound speeds can lead to rather
extreme EOSs, that is, EOSs having rapidly changing material properties,
especially close to where the boundary conditions are imposed. We also
observe this effect in our calculation and find it to be enhanced when
using a large number of segments. In principle, it is possible to
introduce a criterion that discards such solutions with strongly varying
sound speed. For example, \citet{Annala2019} demanded that the energy
densities at two successive inflection points do not have an excessive
relative variation, \ie $(e_i-e_{i+1})/e_i>\Delta \ln e$ with $\Delta \ln
e=0.5$. In this way, models with steep gradients on small energy scales
are filtered out. Here however, and as mentioned in the main text, we
have decided not to apply such a filter but provide evidence in this
section that the first peak in the sound speed is a robust feature that
is independent of the number of segments and filtering.

A sufficiently large number of segments is important to avoid a too
coarse description of the sound speed and related artefacts that we
discuss further below. We note that the setup just described is able to
approximate phase transitions very closely. However, obtaining a
``perfect'' first-order phase transition requires that two consecutive
draws for the sound speed are exactly zero. Since we do not explicitly
demand that this happens, perfect first-order phase transitions are very
unlikely in practice, but our results include a number of cases with very
low sound speed at two neighboring matching points.  On the one hand,
choosing $N$ to be small could lead to EOSs that are rather crude and
possibly suffer from intrinsic biases. On the other hand, if $N$ is very
large, the whole construction can be excessively expensive from a
computational point of view. Hence, an optimal choice needs to be found
for $N$. In Fig.~\ref{plot:cs2seg} we compare the PDFs for different
values of the number of segments with and without filtering. Choosing a
low number of segments (\eg $N=3, 4$) has the effect of influencing the
distribution of the sound speed mostly at intermediate and high densities
close to the perturbative QCD boundary conditions. This is shown in the
first two panels on the left of Fig.~\ref{plot:cs2seg}, which refer to
$N=3, 4$, respectively; the third panel from the left refers instead to
$N=5$ and clearly indicate that a sufficiently large number of segments
is necessary to avoid artefacts at large energy densities. When comparing
the PDF for $N=5$ with that obtained in Fig.~1 of the main text, where
$N=7$, indicates that the latter is already a sufficiently large number
of segments; indeed, experiments carried out on a limited set of EOSs
with $N=9$ has shown that the differences in the PDFs are very
small. Overall, the first three panels of the figure show that the steep
increase of the distribution beyond $c^2_s \simeq 1/3$ at small energy
densities is a robust result that is independent of the number of
segments and filtering. Finally, note that the right-most panel in
Fig.~\ref{plot:cs2seg} refers to a PDF for $N=5$ in which the filtering
on the energy-density gradient is applied. Note how the filtering does
modify the behaviour of the PDF at very large energy densities removing
the most extreme EOSs. It remains unclear whether this should be
considered as more realistic PDF.

\begin{figure}[h!]
\center
\includegraphics[width=0.42\textwidth]{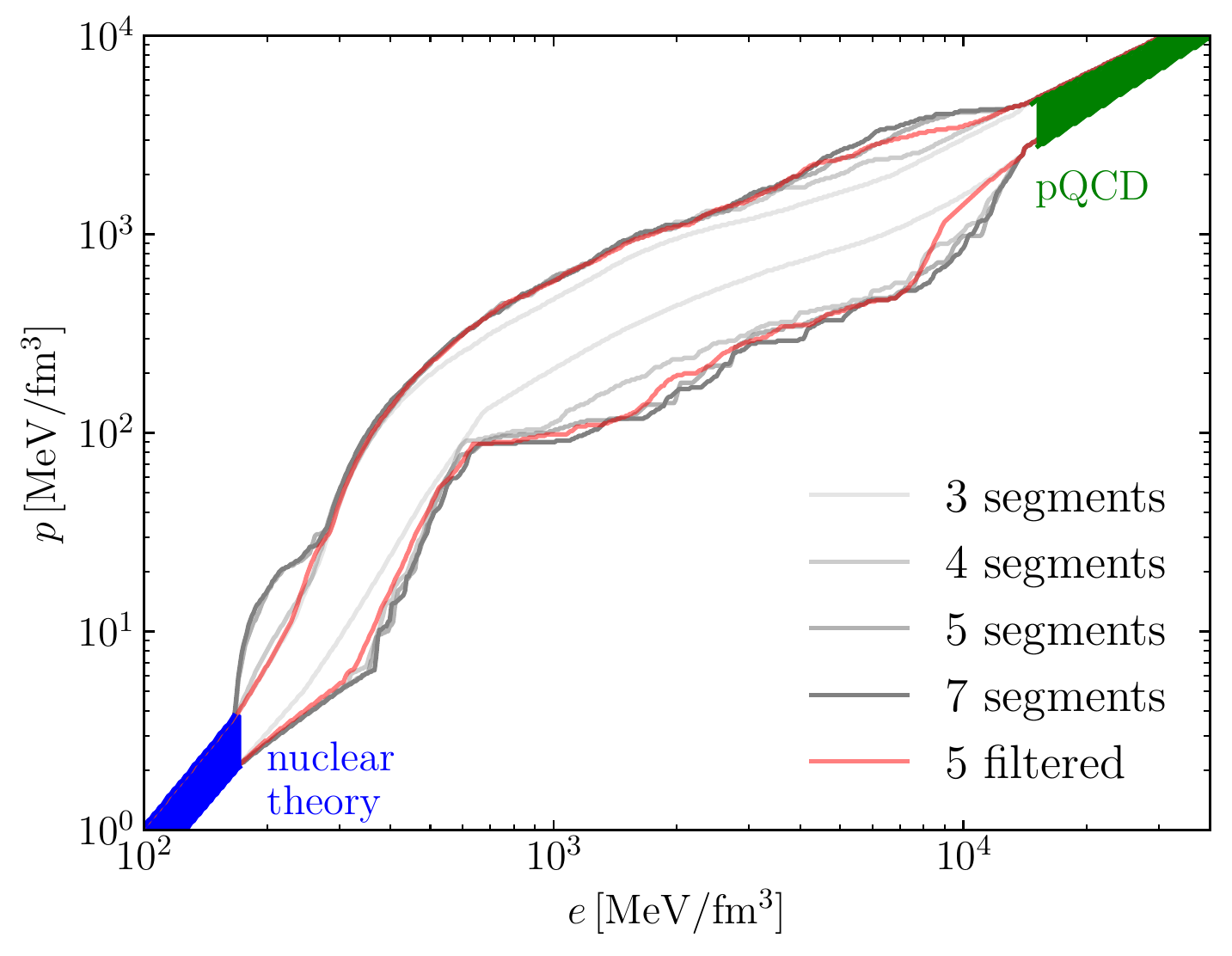}
\caption{Outermost contours of constraint-satisfying EOSs. Gray curves
  are obtained from unfiltered sampling with three, four, five, and seven
  segments; the light red contour refers to the five-segments EOS after
  filtering.}
\label{plot:EOSseg}
\end{figure}

Having a different number of segments also impacts the PDFs of the
pressure as a function of the energy density. This is reported in
Fig.~\ref{plot:EOSseg}, which shows the outer envelope of constraint
satisfying solutions for different numbers of segments and reported with
various gray curves. Note that all PDFs refer to the EOSs that have not
been filtered, while the light-red curve is the result for five segments
and a filter on the relative energy density $\Delta \ln e=0.5$. Note that
a parametrization with three segments results in a band that is
significantly narrower than for larger numbers of segments. However, a
comparison with Fig.~\ref{plot:EOS}, which was done with seven segments,
suggests that most of the missed solutions fall inside the sparsely
populated region. Increasing the number of segments from three to seven
clearly shows that already five segments lead to a well-converged result
for the outer envelope. Finally, inspecting the light-red curve it is
possible to appreciate that the filtering affects mostly the regions
close to the boundary conditions at low and very high energy densities,
leaving the contours in between essentially unchanged.

\begin{figure}[h!]
\center \includegraphics[width=0.42\textwidth]{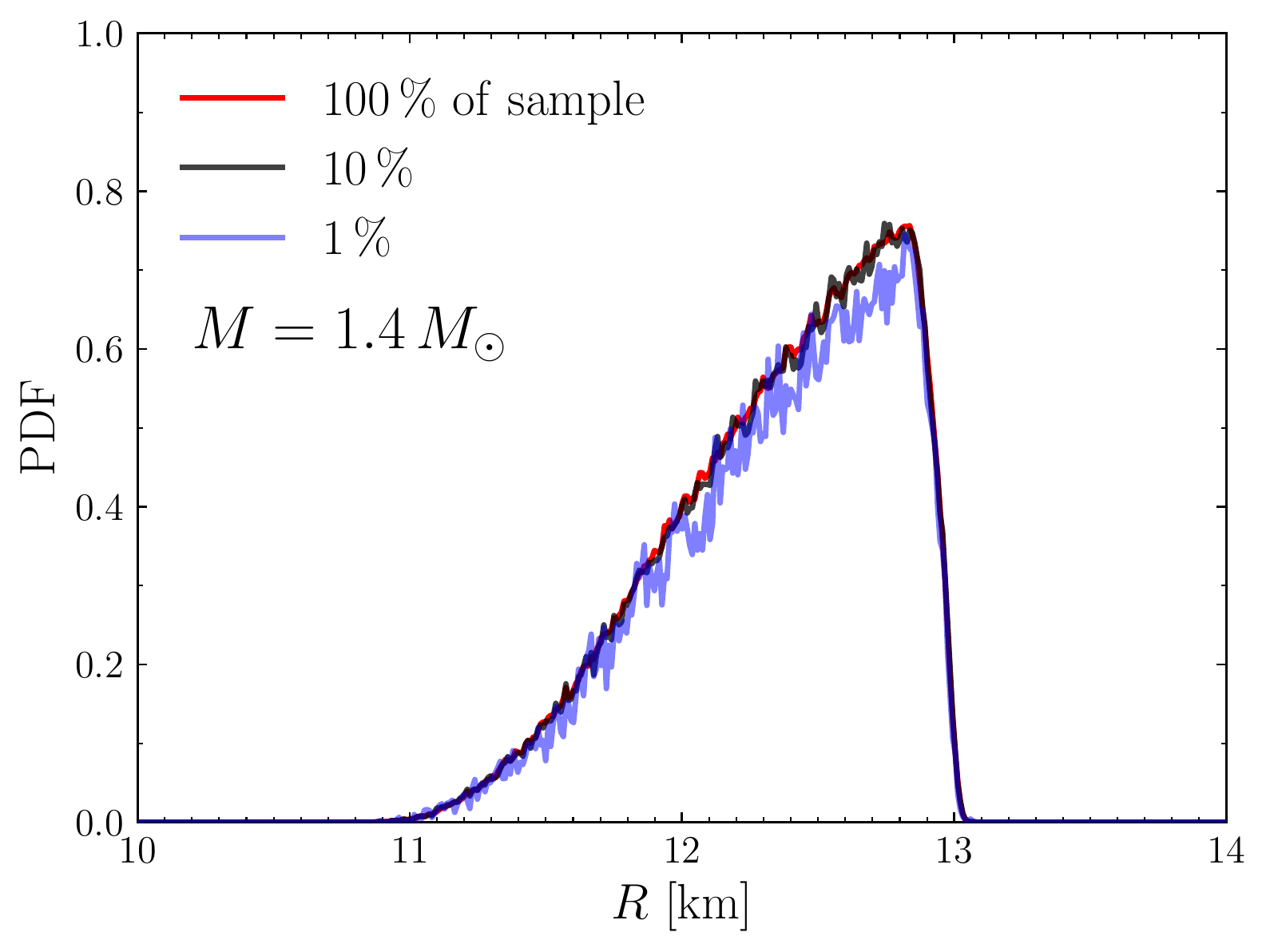}
\caption{PDF of stellar radii at a fixed mass $M=1.4\,M_{\odot}$ for
  different fractions of our sample. Clearly, having a large number of
  EOSs decreases the size of the fluctuations; however $10\%$ of the
  sample is already sufficient to obtain robust estimates of the median.}
\label{plot:stat}
\end{figure}

We should note that the position of the peak in the sound speed at
neutron-star densities is most probably related to the boundary
conditions set at high densities. Ideally, it would be interesting to
study how the properties of the peak in the sound speed varies when
different boundary conditions in the pQCD limit are imposed. Doing so,
however, is far from trivial than it might seems. The actual reason for
this peak \citep[(see also][]{Gorda:2022jvk} is not so much the conformal
value of the sound speed, but, rather, the values of the QCD pressure and
the fact that they must be approached in a causal manner by our sampled
EOS models. In order to relax these boundary condition one would need to
define a new range for the pressure as well as a relation between the
pressure and the energy density that leads to the desired value of the
sound speed. This is equivalent to formulating an alternative EOS model
that would differ from the established perturbative QCD result. If such a
self-consistent model were available and was employed in our analysis, it
would most likely move the location at which the sound speed has to
decrease in order to meet the high-density boundary
conditions. Unfortunately, however, this model is not presently available,
at least to the best of our knowledge; hence, a purely ad-hoc change in
the high-density regime may lead to physically inconsistent results.

Another aspect of our analysis that is worth discussing is the number of
EOS that is necessary to use to have a statistically robust description
of the PDFs. As mentioned in the main text, we have computed at total of
$1.5\times10^7$ EOSs and we have found that such a large number is
sufficient to obtain a variance that is of $\sim3\%$ at most. This is
shown in Fig.~\ref{plot:stat}, which reports a cut of the PDF for a
reference mass of $M=1.4\,M_{\odot}$ when considering different fractions
of our sample. In particular, the red line provides the section of the
PDF when all EOSs are taken into account and corresponds therefore to the
red line reported in the bottom part of Fig.~3 of the main text. Shown
instead with black and blue lines are the cuts when considering only
$10\%$ or $1\%$ of the sample, respectively. Note that having only $10\%$
of the EOSs in the sample is sufficient to obtain robust estimates of
the median.

We conclude this Appendix with a discussion focused on the
properties of the EOSs that lead to sound speeds that are monotonic. As
mentioned in the main text, because these EOSs are rather rare from a
statistical point of view, it is convenient to construct our ensemble
by sampling the sound speed not across all of its possible values, \ie
$0\leq c^2_s \leq 1$, but only in the relevant regime, \ie $0\leq
c^2_s \leq 1/3$. In this way, it was possible to produce a separate and
rich ensemble of $1.5\times10^6$ EOSs with sound speed increasing
monotonically and for which no observational constraints on the maximum
mass were imposed.

\begin{figure*}[ht!]
\includegraphics[width=0.45\textwidth]{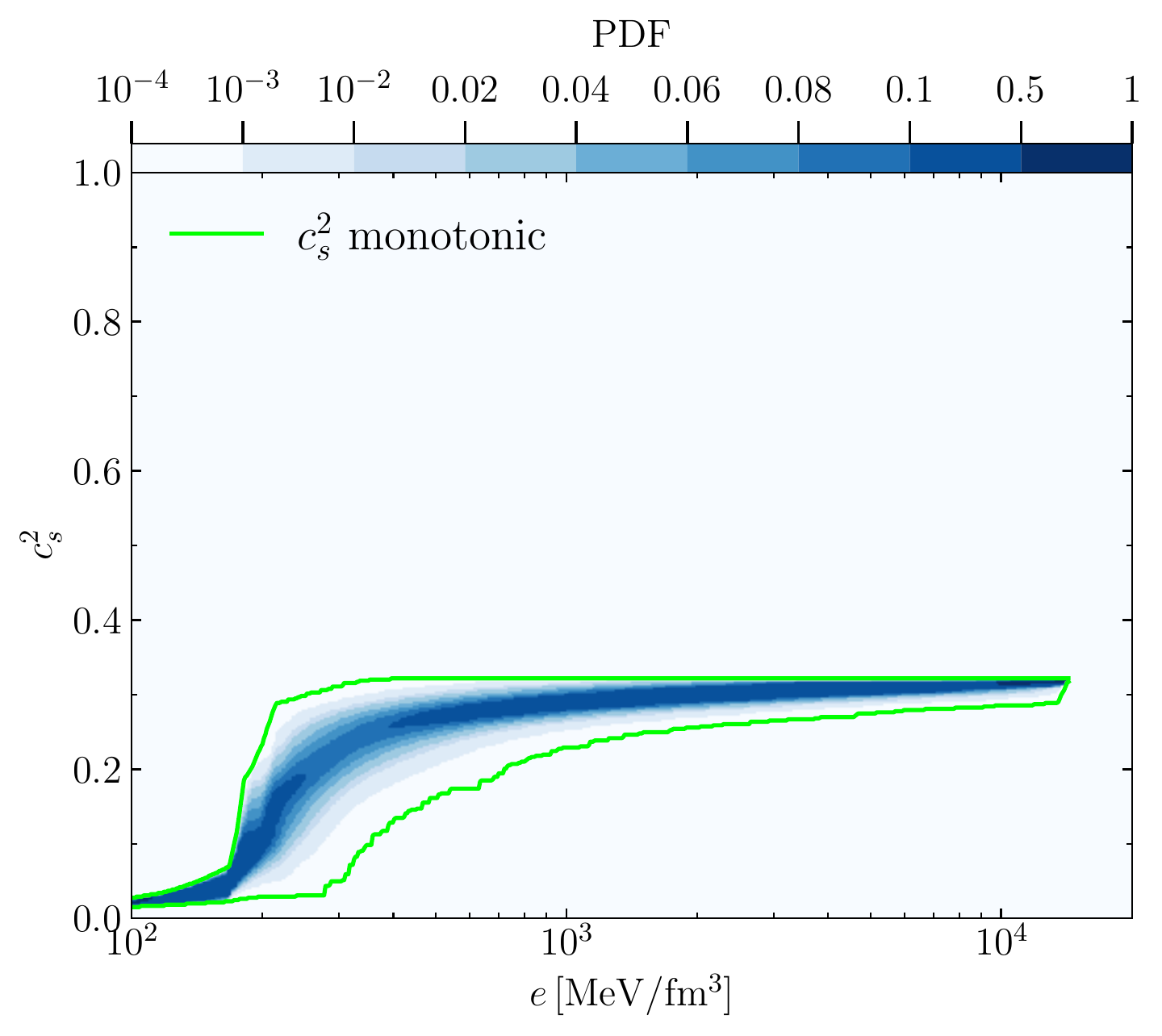}
\includegraphics[width=0.45\textwidth]{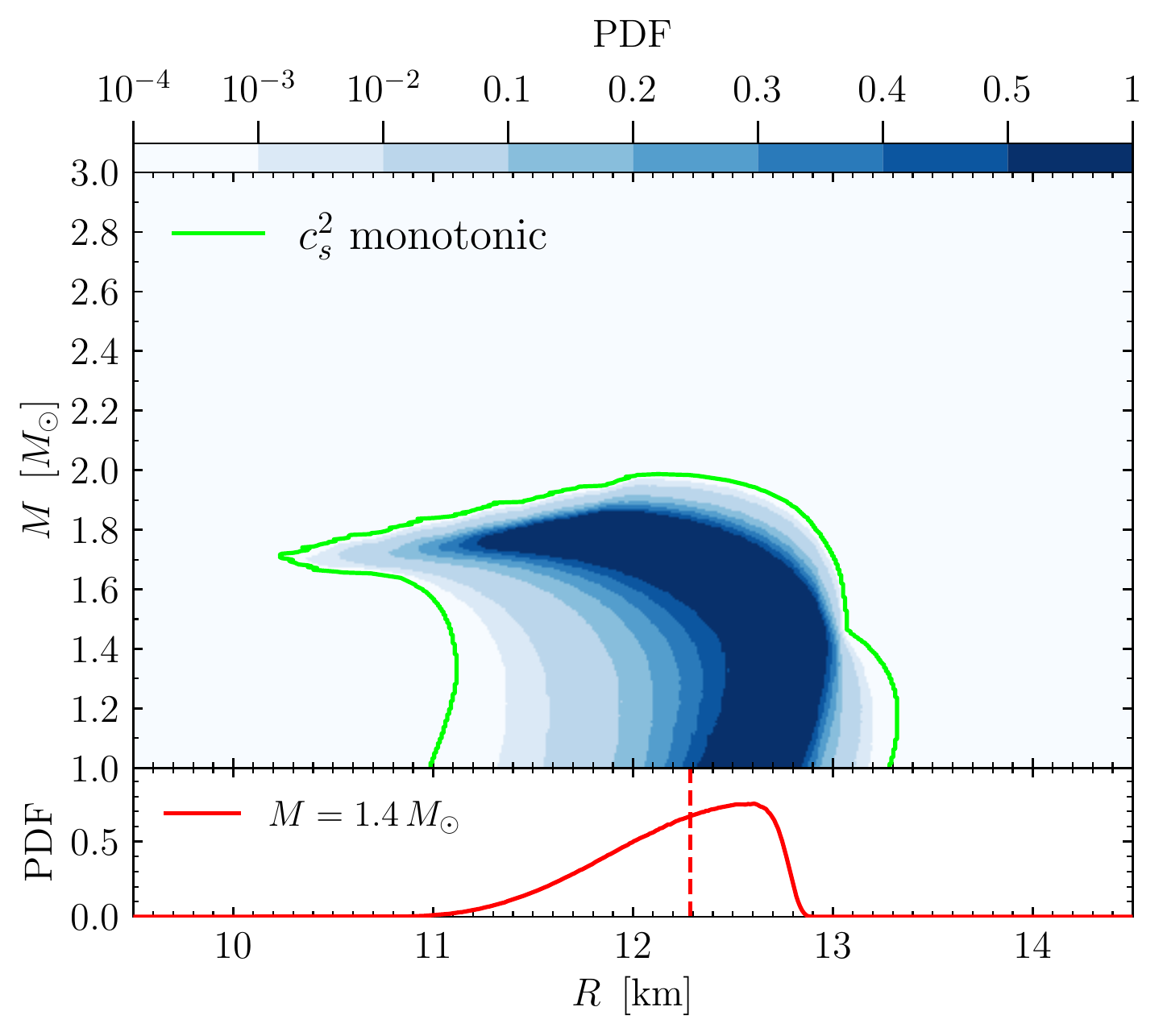}
\caption{PDF for the sound speed (left panel) and the mass-radius relation
  (right right) of monotonic EOSs without observational constraints. Note
  that no stellar model can be found with mass $M \gtrsim 2\,M_{\odot}$.}
\label{plot:MRmono}
\end{figure*}

These results are summarised in Fig.~\ref{plot:MRmono}, whose two panels
show the corresponding PDFs for the sound speed as a function of the
energy density (left panel) and of the mass as a function of the stellar
radius (right panel). In this respect, they can be compared with Figs.~1
and 3 of the main text, where however the green contour is now used to
mark the entire span of the PDFs. 
Interestingly and importantly, the PDFs reveal that the largest mass that
can be sustained in this case is $M\leq1.99M_\odot$.

\newpage
\bibliographystyle{aasjournal}
\bibliography{aeireferences}

\end{document}